\documentstyle[11pt,aaspp4,psfig]{article}
\def\be{\begin{equation}}
\def\ee{\end{equation}}

\def\lsim{\lower 2pt \hbox{$\, \buildrel {\scriptstyle <}\over
         {\scriptstyle \sim}\,$}}
\newcommand\gsim{\buildrel > \over \sim}
\begin{document}
\newcommand{\figureout}[3]{\psfig{figure=#1,width=5.5in,angle=#2} 
   \figcaption{#3} }

\title{Extended Acceleration in Slot Gaps and Pulsar High-Energy Emission}

\author{Alex G. Muslimov\altaffilmark{1} \& 
Alice K. Harding\altaffilmark{2}}   

\altaffiltext{1}{ManTech International Corporation, 
Lexington Park, MD 20653}

\altaffiltext{2}{Laboratory of High Energy Astrophysics, 
NASA/Goddard Space Flight Center, Greenbelt, MD 20771}
 

\begin{abstract}

We revise the physics of primary electron acceleration in 
the ``slot gap" (SG) above the pulsar polar caps (PCs), a regime originally 
proposed by Arons and Scharlemann (1979) in their electrodynamic model of pulsar PCs. 
We employ the standard definition of the SG as a pair-free space between the last 
open field lines and the boundary  of the pair plasma column which is expected to 
develop above the bulk of the PC. The rationale for our revision is that the proper 
treatment of primary acceleration within the pulsar SGs should take into account 
the effect of the narrow geometry of the gap on the electrodynamics within the 
gap and also to include the effect of inertial frame dragging on the particle 
acceleration. We show that the accelerating electric field within the gap, being 
significantly boosted by the effect of frame dragging, becomes reduced because of 
the gap geometry by a factor proportional to the square of the SG width. 
The combination of the effects of frame dragging and geometrical screening 
in the gap region naturally gives rise to a regime of extended acceleration, that 
is not limited to ``favorably curved" field lines as in earlier models, and the 
possibility of multiple-pair production by curvature photons at very high altitudes, 
up to several stellar radii. We present our estimates of the characteristic SG 
thickness across the PC, energetics of primaries accelerated within the gap, 
high-energy bolometric luminosities emitted from the high altitudes in the gaps, 
and maximum heating luminosities produced by positrons returning from the elevated 
pair fronts. The estimated theoretical high-energy luminosities are in good 
agreement with the corresponding empirical relationships for $\gamma $-ray 
pulsars. We illustrate the results of our modeling of the pair cascades and 
$\gamma $-ray emission from the high altitudes in the SG for the Crab pulsar. The 
combination of the frame-dragging field and high-altitude SG emission enables 
both acceleration at the smaller inclination angles and a larger emission beam, 
both necessary to produce widely-spaced double-peaked profiles. 

\end{abstract} 


\pagebreak
  
\section{INTRODUCTION}

There has been a steadily growing number of rotation-powered pulsars with emission detected at high energies, following the success of the Compton Gamma-Ray Observatory (Thompson 2001) and detections by ROSAT, ASCA, RXTE and most recently by Chandra and
XMM (Becker \& Aschenbach 2002).  With the number of pulsars having detected emission above 1 keV 
now approaching several dozen, high-energy radiation seems to be a common feature.  Although 
the theory of acceleration and high-energy emission in pulsars has been studied for over twenty-five years, the origin of the pulsed non-thermal emission is a question that remains unsettled.
The observations to date have not been able to clearly distinguish between an emission site
at the magnetic poles (Daugherty \& Harding 1996, hereafter DH96) and emission from the outer magnetospheric gaps
(Cheng, Ho \& Ruderman 1986, Hirotani \& Shibata 2001).  Complicating the problem is the fact that both models 
have some unresolved difficulties.  In the case of polar cap (PC) models, while the energetics and pair-cascade spectrum have had success in reproducing the observations, the predicted beam size of 
radiation emitted near the neutron star (NS) surface is too 
small to produce the wide pulse profiles 
that are observed.  However, Arons (1983, hereafter A83) first noted the possibility of a high-altitude acceleration region or ``slot gap" (SG) near the PC rim, based on the finding of Arons \& Scharlemann (1979, hereafter 
AS79) that the pair formation front (PFF), above which the accelerating field is screened, occurs at increasingly higher altitude as the magnetic colatitude approaches the last open field line
where the electric field vanishes. The SG is a characteristic feature that is unavoidable in any PC space-charge limited  flow model. 
In the last few years, a self-consistent model of PC acceleration  
has been developed, which includes effects of general relativity (Harding \& Muslimov 1998, hereafter HM98) 
and screening by pairs (Harding \& Muslimov 2001, hereafter HM01; Harding \& Muslimov 2002, hereafter HM02).  HM98 concentrated on acceleration in the
interior regions of the PC, but their calculation confirmed that a SG also forms near the PC rim in the presence of inertial frame dragging. Interestingly, since the magnitude, and longitudinal and radial dependences of the accelerating electric field used in HM98 are different from those in the model of AS79, in the calculation of HM98 the PFF is curving up more abruptly to form the inner boundary of a much narrower SG. 

The present paper is motivated by the fact that, despite the attractive possibility of primary acceleration within the SG region of a pulsar inner magnetosphere (AS79), the electrodynamics and energetics of the 
SG acceleration remain to a large extent unexplored beyond the original model of AS79. What we know about SG acceleration is mostly based on a classical model of AS79 and further developments by A83 and 
Arons (1996, hereafter A96), and it has never been investigated in any other electrodynamic model thus far. However, we believe that within the general-relativistic model of space-charge limited flow from the pulsar PC (Muslimov \& Tsygan, 1992, hereafter MT92) the acceleration properties within the SG will significantly differ from those implied by the SG originally proposed by AS79. 
The SG model of AS79 and A83 produced effective acceleration only for large inclination
angles and only on ``favorably curved" field lines, i.e. those field lines curving
toward the rotation axis.  The frame-dragging acceleration dominates at most pulsar periods at small inclination angles, and operates over the entire PC (i.e. it is not
restricted to favorably curved field lines).
Besides this simple update of the underlying electrodynamic model, we incorporate an
additional effect of screening of the electric field within the PC slot, a narrow space between the last open field lines and lines confining the pair plasma column generated at 
relatively low altitudes above the PC surface. This enhanced screening is caused by 
taking into account a second conducting boundary formed by the pair plasma column very near the last open field line.  The $E_{\parallel}$ within the SG is greatly reduced by
the proximity of two conducting boundaries and is 
well-known from the electrodynamics in narrow tubes and cones with conducting surfaces (Landau \& Lifshitz 1984).  Ironically, this effect of screening by two closely spaced conducting boundaries was overlooked in original calculations of the SG acceleration, in which only the conducting boundary at the last open field line was taken into account. To estimate the field in the SG, A83 simply applied the solution for the accelerating 
electric field in the absence of the pair plasma, assuming screening only from the last 
open field line boundary.

The major advantage of acceleration within the SG is the possibility of establishing the curvature radiation (CR) PFF and generation of $\gamma -$ray emission at high altitudes. The need to locate the generation of pair cascades and $\gamma$-ray emission
at high altitude above the PC was demonstrated by Daugherty \& Harding (1996, hereafter DH96), and later also by Rudak et al. (2002) who numerically modeled the pair formation, high-energy radiation spectra and light curves at different altitudes in pulsars. They found that the shape of the cascade spectrum and high-energy cutoff, due to magnetic
pair creation, for the Vela pulsar best matches the observed spectrum for emission at 
altitudes around 1-2 stellar radii above the surface. Furthermore, the wide $\gamma$-ray
profiles observed can be produced in PC models only if the emission beam opening angle
is comparable to the inclination angle.  High altitude emission along flaring field lines
would enable larger high energy beams.
DH96 approached the problem by trying to reproduce the observed spectra and light curves of the Vela pulsar by assuming a 
simple parameterized model of the accelerating electric field.  However, the physical origin of that accelerating field as well as possible 
regimes of acceleration and energetics for different pulsar spin periods and magnetic fields could not be addressed in that approach. Also the details of such modeling, such as the pair-formation altitudes, 
hardness of the spectrum, and widths of the high-energy pulses, will also depend on the regime of acceleration of primaries within the SG.  
In his study of high-energy emission from the SG, A96 came to the conclusion that the SG energetics, as determined by a potential which did not include the frame-dragging effect, is not enough to account 
for the observed $\gamma -$ray luminosities and spectra of the Crab and Vela pulsars.  Furthermore,
he investigated only the CR from the primary particles accelerating in the SG which
results in a spectrum that does not match the softer observed spectra of $\gamma$-ray pulsars.
In this paper, we additionally consider the emission from the high-altitude pair cascades above 
the PFF that bounds the inner edge of the SG.  We derive the general relativistic accelerating electric field 
in the case of the SG to include the effect of screening 
caused by the slot geometry, and present our calculations of pair cascades at high altitudes (at a few 
stellar radii above the surface), $\gamma -$ ray spectra, and $\gamma -$ ray light curves. We provide 
simple expressions for the $\gamma $-ray and heating luminosities, and discuss possible constraints on the Crab pulsar viewing angles and obliquities.    
       
The paper is organized as follows. In \S 2 we discuss the electrodynamics 
within the SG regions of pulsars. We present the electric potential 
drop in \S 2.1, calculate the energetics of primaries in \S 2.2, estimate 
the characteristic gap thickness in \S 2.3, and derive $\gamma -$ray 
luminosities per solid angle emitted by the SGs as well as heating 
luminosities in \S 2.4. In \S 3 we discuss the main results of our numerical cascade simulations. Finally, in \S 4 we discuss our main results and draw our principal conclusions.

\section{Electrodynamics of Slot Gap}

The SG model for the emission region of a pulsar was originally 
proposed by AS79, who were the first to observe 
that in a space-charge limited flow the process of pair formation may significantly vary across the PC. For example, in their electrodynamic model, the pair formation surface tends to curve up and become asymptotically tangent to the field lines near the boundary of the 
polar tube in its half with the so-called ``favorably curved" field lines. This occurs simply because the acceleration of electrons in this region is not sufficient to produce pairs at low altitude, leaving a SG  
around the column of pair plasma. This model was described 
in detail by A83, and also overviewed by Arons (1981) and A96. The 
very concept of SG acceleration, and its potential impact on pair production, radio and high-energy emission from pulsars, was a significant finding. However, the original SG treatment failed to produce a viable model of pulsar 
high-energy emission.  One reason is that the calculation of electron acceleration within 
the SG was inconsistent with the acceleration and pair formation in rest of polar flux 
tube region. In this paper we discuss and calculate the effect that we believe makes the 
regime of SG acceleration very feasible and attractive.   

Figure 1 schematically illustrates the geometry of the open field lines at the PC. The extended SG 
region is squeezed between the pair plasma column and closed magnetospheric region filled with plasma. The 
ring-like structure on the top indicates the elevated pair formation site. 

\subsection{Electric Potential Drop within the Slot Gap}

The SG region is bounded by the last open field lines and the lines with the magnetic colatitude 
$(1-\Delta \xi _{\sc SG})$ times smaller, where $\Delta \xi _{\sc SG}$ is the latitudinal 
gap thickness in units of $\xi $ ($\xi = \theta /\theta _0 $ is dimensionless colatitude of a PC field line, and 
$\theta _0 \approx (\Omega R /cf(1))^{1/2}$ is the footpoint of the last open field line).  
Let us introduce the field line separating the SG into two halves and having the magnetic colatitude 
$\theta _{0,\sc SG}= \theta _0 (1-\Delta \xi _{\sc SG}/2)$, and denote the innermost half-space of a SG 
as region I and the outermost half-space as region II. Then the magnetic field lines within each of these regions 
can be described by the colatitudes
\begin{eqnarray}
\theta _{\sc SG}(\eta ) & = &\theta _{0, \sc SG} \left[ \eta {f(1)\over f(\eta )} \right]^{1/2} \left[ 1-{1\over 2}\Delta \xi _{\sc SG}\xi _{\ast }\right] = \nonumber \\ 
& & \theta _0 \left[ \eta {f(1)\over f(\eta )} \right]^{1/2} \left[ 1-{1\over 2}\Delta \xi _{\sc SG} (1+\xi _{\ast })\right], ~~~~~ \xi _{\ast } \in [1,0]~~~~~{\rm for~~region~~I}
\label{thetaI}
\end{eqnarray}
and 
\begin{eqnarray}
\theta _{\sc SG}(\eta ) & = & \theta _{0, \sc SG} \left[ \eta {f(1)\over f(\eta )} \right]^{1/2} \left[ 1+{1\over 2}\Delta \xi _{\sc SG}\xi _{\ast }\right] = \nonumber \\
& & \theta _0 \left[ \eta {f(1)\over f(\eta )} \right]^{1/2} \left[ 1-{1\over 2}\Delta \xi _{\sc SG} (1-\xi _{\ast })\right],
~~~~~ \xi _{\ast } \in [0,1]~~~~~{\rm for~~region~~II,}
\label{thetaII}   
\end{eqnarray}
respectively.\\
Here $\xi _{\ast }$ is the dimensionless colatitude of field lines 
bounded by the SG regions I and II and which is counted from the 
field line that separates those regions, $\eta = r/R$ is the dimensionless radial coordinate, and $f(\eta )$ is the correction factor 
for the dipole component of the magnetic field in a Schwarzchild metric (see MT92 for details). 

We shall also introduce the SG half-width in the longitudinal direction scaled to the PC surface, 
$\delta _{\sc SG} = (1/2) r_{\rm pc}\Delta \xi _{\sc SG}$, where $r_{\rm pc}=R\theta _0$ is the PC radius.        
In a small-angle approximation (which is perfect for the SG case) the Goldreich-Julian (GJ) charge 
density and the charge density of primary electrons read  
\be
\rho _{_{\sc GJ}} = - {{\Omega B_0}\over {2\pi c \alpha \eta ^3 }} {{f(\eta )}\over {f(1)}}
\left[ (1-{\kappa \over {\eta ^3 }})\cos \chi + {3\over 2} \theta _{\sc SG}(\eta ) H(\eta )\sin \chi \cos \phi _{\rm pc}\right]  
\label{rhoGJ}
\ee
and 
\be
\rho  = - {{\Omega B_0}\over {2\pi c \alpha \eta ^3 }} {{f(\eta )}\over {f(1)}}
\left[ (1-\kappa )\cos \chi + {3\over 2} \theta _{\sc SG}(1) H(1) \sin \chi \cos \phi _{\rm pc} \right] , 
\label{rho}
\ee
respectively. \\ 
Here $\Omega = 2\pi /P$, $P$ is the pulsar spin period, $B_0$ is the surface value of the magnetic field strength at magnetic pole, $\alpha $ is the redshift function, $\kappa = (r_{\rm g}/R) (I/I_0) \approx 0.15~I_{45}/R_6^3$ is the general relativistic parameter entering the frame-dragging effect, $r_{\rm g}$ is the NS gravitational radius, $I_{45} = I /10^{45}$ 
g$\cdot $cm$^2$, $R_6 = R/10^6$ cm , $I$ is the NS moment of inertia, $I_0 = MR^2$, $M$ and $R$ are, respectively,  NS mass and radius, $H$ is a
relativistic correction factor of order one, $\chi $ is the 
pulsar obliquity, and $\phi _{\rm pc}$ is the magnetic azimuthal angle (see MT92 for 
details).

For high enough altitudes within the SG, i.e. for $\eta -1 \gg \delta _{\sc SG}$, and in a small-angle approximation,  
the general-relativistic Poisson's equation (see also the derivation of eq. [37] in MT92),
\be
\nabla \cdot \left( {1\over \alpha } \nabla \Phi \right) = - 4 \pi (\rho - \rho _{_{\sc GJ}}),
\ee
translates into
\be
{1\over {\alpha \theta \eta ^2}}
\left[ {{\partial }\over {\partial \theta }}\left( \theta  
{\partial \over {\partial \theta }}\right) + {1\over {\theta }} {{\partial ^2}
\over {\partial \phi _{\rm pc}^2}} \right] 
\Phi (\eta,\theta, \phi _{\rm pc}) 
= - 2 {1\over {\alpha \eta ^3}}{{f(\eta )} \over {f(1)}} 
\Phi _0 ( {\cal A} \cos \chi + {\cal B} \sin \chi \cos \phi _{\rm pc}),
\label{poisson1}
\ee
where
\be
{\cal A} = \kappa (1-1/\eta ^3),
\ee
\be
{\cal B} = (3/2)[H(\eta )\theta _{\sc SG}(\eta ) - H(1)\theta _{\sc SG}(1)].
\ee
By employing expressions (\ref{thetaI}), (\ref{thetaII}) and changing variables from $\theta $ to $\xi _{\ast }$ in eq. (\ref{poisson1}), and also assuming the approximation $\Delta \xi _{_{\sc SG}}\ll 1$ we get
\be
\left( {1\over {\nu _{_{\sc SG}}}} {{\partial ^2}\over {\partial \xi _{\ast }^2}} 
+ {{\partial ^2}\over {\partial \phi _{\rm pc}^2}} \right) 
\Phi (\eta, \xi _{\ast },\phi _{\rm pc}) = 
- 2~\Phi _0~\theta _0^2~({\cal A} \cos \chi + {\cal B} \sin \chi \cos \phi _{\rm pc}),
\label{poisson}
\ee
where 
\be
\nu _{_{\sc SG}} = {1\over 4} \Delta \xi _{_{\sc SG}}^2 =  
\left( {{\delta _{_{\sc SG}}}\over {r_{\rm pc}}} \right) ^2 
\label{nu}
\ee
is the screening factor originating from the gap geometry. Also in eq. (\ref{poisson})  
$\Phi _0 = (\Omega R/c) B_0~R$, which is the characteristic value of the maximum potential drop in a vacuum solution (see Deutsch 1955). 

Assuming that the SG surfaces are equipotential ($\Phi [\xi _{\ast }=1]=0$), the solution of (\ref{poisson}) 
for the electric potential within regions I and II of a SG can be written as  
\be
\Phi (\eta, \xi _{\ast },\phi _{\rm pc}) 
= \Phi _0  \theta _0^2 \left\{ \nu _{_{\sc SG}} {\cal A} (1-\xi _{\ast }^2)\cos \chi + 2~{\cal B}~\left[ 1- {{\cosh (\sqrt{\nu_{_{\sc SG}}}\xi _{\ast })}\over {\cosh(\sqrt{\nu _{_{\sc SG}}})}} \right] \sin \chi \cos \phi _{\rm pc } \right\} .
\label{Phi1}
\ee
Since $\nu _{_{\sc SG}} \ll 1$, we can write that $2[1-\cosh(\sqrt{\nu _{_{\sc SG}}}\xi _{\ast })/\cosh(\sqrt{\nu _{_{\sc SG}}})] \approx \nu _{_{\sc SG}}(1-\xi _{\ast }^2)$. Hence, 
eq. (\ref{Phi1}),  after substitution the explicit expressions for ${\cal A}$ 
and ${\cal B}$, translates into 
\begin{eqnarray}
\Phi (\eta, \xi _{\ast },\phi _{\rm pc}) 
& = & \Phi _0  \theta _0^2 \nu _{_{\sc SG}} \left[ 
\kappa \left( 1 - {1\over \eta ^3} \right) \cos \chi + {3\over 2}~\theta _{0,\sc SG} H(1) \times \right. \nonumber \\
& & \left. \left( {{H(\eta )}\over {H(1)}} \sqrt {\eta {f(1)\over f(\eta )}} - 1 \right) \sin \chi \cos \phi _{\rm pc} \right] (1-\xi _{\ast }^2) .
\label{Phi}
\end{eqnarray} 
Note that the angular size of the SG limits the maximum lengthscale of the 
angular part of Laplacian and naturally enters the solution of Poisson's equation (see MT92 for details). The factor $\nu _{_{\sc SG}}$, to be evaluated in Section 
\ref{sec_SGthick}, is proportional to the square of the SG thickness $\Delta \xi _{\sc SG}$.  Thus, the additional screening introduced by the conducting boundary at the
PFF results in a reduction of the potential $\Phi(\eta)$ and the associated 
$E_{\parallel}$ in the SG.  It is important that from the 
electrodynamic point of view, we treat the SG in exact same way we generally treat the whole region of 
open field lines of the PC. Thus, the physics of screening effect within the SG is akin to the physics of screening of the vacuum electric field within 
the open field line region of pulsar by the geometry of the region itself.

\subsection{Primary Energetics within the SG and Emerging $\gamma $ - Ray Flux}

Let us write the general formula for the luminosity of primaries accelerating in a SG,
\be
L_{\rm prim} (\eta _{_{\sc PFF}}) = \alpha c 
\int _{_{{\cal S}_{_{\sc SG}}(\eta _{_{\sc PFF}}) }}|\rho (\eta ) | \Phi (\eta ) dS (\eta ),
\label{Lprim}
\ee
where the integration is over the surface cut in a sphere at the 
radial distance $\eta _{_{\sc PFF}}$ by the SG boundaries, and 
$\eta _{_{\sc PFF}}$ refers to the elevated PFF within the SG region.

In eq. (\ref{Lprim}) $dS$ is the element of a spherical surface at radial distance $\eta $. 
After inserting expressions (\ref{rho}) and (\ref{Phi}) into eq. (\ref{Lprim}), we get 
\be
L_{\rm prim} = f_{_{\sc SG}} L_{\rm sd},
\label{Lprim2}
\ee
where
\begin{eqnarray}
f_{_{\sc SG}} & = & \nu _{_{\sc SG}} \Delta \xi _{_{\sc SG}} 
\left\{ \kappa (1-\kappa)\left( 1- {1\over \eta ^3} \right) \cos ^2 \chi +
\right. \nonumber\\ 
& & \left. {9\over 8} \theta _0^2 H^2(1)\left[ {H(\eta )\over H(1)} \sqrt {\eta {f(1)\over f(\eta )}} -1 \right] 
\sin ^2 \chi \right\} _{|_{\eta = \eta _{_{\sc PFF}}}} ,
\label{fSG}
\end{eqnarray}
is the efficiency of converting the spin-down power into the power of primary particles within the SG, and $L_{\rm sd} = \Omega ^4 
B_0^2 R^6/6c^3f^2(1)$ is the pulsar spin-down luminosity, 
where $B_0/f(1)$ is the surface value of the magnetic field 
strength corrected for the ``gravitational redshift". 
 
Note that the efficiency $f_{_{\sc SG}}$ is a function of pulsar 
parameters $B$ and $P$ (via $B,~P$ - dependence of  $\nu _{_{\sc SG}}$ and 
$\Delta \xi _{_{\sc SG}}$), so the luminosity of a primary beam scales as  
$L_{\rm prim}\propto L_{\rm sd}^{\alpha }$, where $\alpha = 1/7$ and 5/14  
for $B \lsim 0.1~B_{\rm cr}$ and $\gsim 0.1~B_{\rm cr}$, respectively 
(see \S ~2.4 for details).  

\subsection{Estimate of Slot Gap Longitudinal Thickness} \label{sec_SGthick}

In this paragraph we estimate the characteristic scale of a SG  
over the colatitude $\xi $-coordinate. We shall discriminate between two cases where the local magnetic field
$B \lsim 0.1~B_{\rm cr}$ and $B \gsim 0.1~B_{\rm cr}$, respectively, where 
$B_{\rm cr} = 4.413\cdot 10^{13}$ G is the quantizing value of the magnetic 
field strength. For $B < 0.1~B_{\rm cr }$, the attenuation length of the CR photons is 
determined by the condition $\epsilon B'\sin \theta _{\rm kB} \gsim 0.2$, whereas 
for $B > 0.1~B_{\rm cr}$ it is determined by the threshold condition 
$\epsilon \sin \theta _{\rm kB} > 2$, where $\epsilon $ is the photon energy in 
units of $m_{\rm e}c^2$, $B'=B/B_{\rm cr}$, and $\theta _{\rm kB}$ is the angle between the photon momentum and 
tangent to the magnetic field line. The explicit formulae for the attenuation length read (Harding \& Daugherty 1983, Harding et al. 1997)
\be
s = {{0.2~\rho _{\rm c}}\over {B'\epsilon }},~~~~~{\rm if}~~~~B \leq 0.1~B_{\rm cr}  
\label{atten1}
\ee
and 
\be
s = {{2\rho _{\rm c}}\over \epsilon}, ~~~~~{\rm if}~~~~~B \geq 0.1~B_{\rm cr},
\label{atten2}
\ee
where $\rho _{\rm c} \approx (4/3)\sqrt{cr/\Omega}$ is the radius of curvature. 

\noindent
{\it The case $B \lsim 0.1~B_{\rm cr}$}

Let us estimate the characteristic thickness of the SG, $\Delta \xi _{_{\sc SG}}$, for
the case $\chi \approx 0$. Since the acceleration 
within the SG occurs mostly in the saturation regime (i.e. at high altitudes, where the accelerating 
$E_{\parallel}$ saturates), we can use the following expression for the dimensionless altitude $z_0$ 
(here in units of stellar radius) of pair formation due to CR (cf. eq. [40], for the saturated regime, in HM01) as a function of $\xi $, to estimate the characteristic scale of a SG over $\xi $, 
\be 
z_0 = 7 \times 10^{-2}~{{P_{0.1}^{7/4}}\over {B_{12}~I_{45}^{3/4}}} 
{1\over {\xi ^{1/2} (1-\xi ^2)^{3/4}}},
\label{zeta1}
\ee
where $P_{0.1} = P/0.1$ s, $B_{12} = B_0/10^{12}$ G.  In our derivation of formula
(\ref{zeta1}) we employed the same procedure of minimization of the sum of acceleration
length and photon mean-free path, as we did in HM98, HM01 and HM02, with the 
explicit $\xi$ dependence of radius of curvature and accelerating electric field 
(for $\chi \approx 0$) included.
The PFF is typically relatively flat over the bulk of the PC 
(for $0.2 \lsim \xi \lsim 0.8$), and begins turning up as $\xi $ 
approaches the rim of the open field line domain and also the 
magnetic axis. Near the rim the PFF tends to establish at a few stellar radii, so that the colatitude 
$\xi _{_{\sc SG}}$ of the beginning of the SG may be estimated via the condition
\be 
{{\partial z_0}\over {\partial \xi }}_{|_{\xi = \xi _{_{\sc SG}}}} 
\sim \lambda ,
\label{dzeta1}
\ee
where  $\lambda \sim 0.1-0.5$ is a free parameter that can be constrained for the 
known $\gamma -$ray pulsars by comparison their observed and predicted $\gamma -$ray 
luminosities. The condition (\ref{dzeta1}) 
means that the regime of the SG acceleration sets in when the  characteristic 
variation of the PFF altitude with $\xi $-coordinate 
rescaled by the whole interval of PC $\xi $ values ($\sim 1$) becomes comparable to $\lambda $ 
times the stellar radius. Note that formula (\ref{dzeta1}) does not take into account the 
effect of screening near the SG and is based therefore on a somewhat 
underestimated value of $z_0$ which results in the underestimation 
of the SG width. For this reason the most likely value of the
parameter $\lambda $ is expected to be in the lower range, say 
$\lambda \sim 0.1-0.3$. 
Earlier (see HM98 and HM01) 
we calculated the shape of the PFF produced at low altitudes by CR 
photons. In general, those calculations are valid only for the   values 
of $\xi $ where the PFF is relatively flat, because they did not take into account 
the effect of electric field screening under consideration in the SG region (which will 
increase the PFF heights there). 
However, formally, we can extend 
our calculations for the values of $\xi $ close to 1, so that we can 
see how the PFF turns up as we approach the rim. Figures 2a and 2b show 
the height of the PFF produced by CR photons at low altitudes as a function of 
$\xi$ near the rim region. We present calculations for the values of 
pulsar spin period $P = 0.1$ s (Figure 2a) and 0.2 s (Figure 2b) and for 
the discrete values of B, ranging from $0.01~B_{\rm cr}$ through 
$0.2~B_{\rm cr}$. Even though our calculations shown in Figures 2a and 2b 
are not consistent in the region where the PFF is curving up, they may provide a rough 
idea about the characteristic width of the SG region. Figures 2a,b show that  
the typical values of $\xi $ at which the heights of PFF begin to 
rapidly grow roughly agree with our estimates of the SG 
width (see eqs. [\ref{deltaxi1}] and [\ref{deltaxi2}] below).    

Assuming that $\Delta \xi _{_{\sc SG}} \lsim 0.3$, from eqs (\ref{zeta1}) and (\ref{dzeta1}) we find that 
\be
\Delta \xi _{_{\sc SG}} \approx 0.2~P_{0.1}(\lambda B_{12})^{-4/7}I_{45}^{-3/7}.
\label{deltaxi1}
\ee

\noindent
{\it The case $B \gsim 0.1~B_{\rm cr}$}

First of all, note that this case was not treated by HM01, but in fact may 
be important for 
high-energy pulsars. Using the threshold condition for the pair attenuation length 
(see eq. [\ref{atten2}]), we can write
\be
z_0 = 5\times 10^{-2}~ {{P_{0.1}^{7/4}} \over {B_{12}^{3/4}I_{45}^{3/4}}} {1\over {\xi ^{1/2} (1-\xi ^2)^{3/4}}}.
\ee
so that for the condition (\ref{dzeta1}) we arrive at the estimate,
\be
\Delta \xi _{_{\sc SG}} \approx 0.1~P_{0.1}(\lambda B_{12}^{3/4})^{-4/7}
I_{45}^{-3/7}.
\label{deltaxi2}
\ee
Note that $\Delta \xi _{_{\sc SG}}$ tends to increase with 
decreasing $\lambda$. 

By substituting expressions (\ref{deltaxi1}) and (\ref{deltaxi2}) into  
formula (\ref{Phi}), we can estimate the characteristic electric potential 
drop as 
\be 
\Delta \Phi \approx 1.5\times 10^{12}~[1-{3\over 8}(1+7\delta)]~B_{12}^{\delta } \left({{I_{45}}\over 
{\lambda ^8}}\right)^{1/7}{\cal F}~~~~~~{\rm Volts},
\label{Phi_last}
\ee
where $\delta = -1/7$ and $1/7$ for the low-B and high-B case, respectively, and 
\begin{eqnarray}
{\cal F} & = & \left[ \kappa_{0.15}\left(1-{1\over {\eta ^3}} \right) \cos \chi + \right. \nonumber\\ 
& & \left. 3.8\times 10^{-3} \left( {R_6^7 \over {P_{0.1}~I_{45}^2}}\right) ^{1/2}\left( H(\eta )\sqrt{\eta {{f(1)}\over{f(\eta )}}}-H(1)\right)
\sin \chi \cos \phi _{\rm pc} \right] (1-\xi _{\ast }^2).
\end{eqnarray}
The above formula for $\Delta \Phi $ illustrates that the accelerating potential (the Lorentz 
factor of primary electrons) is only weakly dependent on the pulsar magnetic field,  
and the dependence on pulsar spin period enters only the second term in expression 
for ${\cal F}$ which is proportional to $\sin \chi $. The strong dependence of the parallel
electric field on magnetic field and spin period have been mostly cancelled by the screening
parameter $\Delta \xi _{_{\sc SG}}$ which depends on the pair production mean-free path. 
The main factors determining 
$\Delta \Phi $ are thus, the radius and  moment of inertia of a NS, and pulsar obliquity 
angle $\chi $ (note that $\kappa _{0.15} = I_{45}/R_6^3$).   

By substituting expressions (\ref{deltaxi1}) and (\ref{deltaxi2}) 
into formula (\ref{Phi_last}), we can now estimate  
the characteristic Lorentz factor of primary electrons, 
$\gamma = e\Delta \Phi /m_{\rm e}c^2$, 
\be
\gamma \approx 4\times 10^7~\left( {{I_{45}}\over {\lambda _{0.1}^8}}
\right) ^{1/7}~[1-{3\over 8}(1+7\delta)]~B_{12}^{\delta } 
\left( 1-{1\over {\eta _{\rm acc}^3}} \right),
\label{gamma}
\ee
where $\eta _{\rm acc}$ is the value of 
$\eta $ corresponding to the characteristic acceleration altitude, 
and $\lambda _{0.1} = \lambda /0.1$.  
For the parameters of a Crab-like pulsar, $B_{12}=5$, $I_{45}=4$, 
and assuming $\eta _{\rm acc} = 2$, and 
$\delta = -1/7$ (low-B case), we get
\be
\gamma \sim 3\times 10^7~\lambda _{0.1}^{-8/7}.
\ee 

In the present analysis the width of the SG in units of dimensionless 
coordinate $\xi $ is constant as a function 
of altitude. In reality, it may slightly vary with  
altitude. For example, it may grow with altitude because 
the plasma column boundary may be getting less sharp with altitude, so that the efficiency of geometrical screening may slightly degrade (meaning the increase in $\Delta \nu _{_{\sc SG}}$) with altitude. Loosely speaking, the effect of fuzziness of the plasma column boundary at high altitudes can 
be attributed to the systematic increase in the deviation of the transverse 
component of the photon mean-free path from the field line toward the 
magnetic axis.

\subsection{The Gamma-Ray Luminosity per Unit Solid Angle Emitted by the SGs}

Here we calculate the specific high-energy luminosity, 
$L_{\gamma }(\Omega _{\gamma })$, radiated by particles 
accelerated within the SGs and emitted into the solid angle 
$\Omega _{\gamma }$. 

First, let us calculate the solid angle within which the photons produced by SGs are emitted,
\be
\Omega _{\gamma } = 2 \pi 
\int _{_{\theta _{{\rm min}, \gamma }^{\sc SG} }}
^{\theta _{{\rm max}, \gamma }^{\sc SG}} 
\sin \theta d\theta ,  
\label{omega1}
\ee
where 
\be
\theta _{{\rm min}, \gamma }^{\sc SG} \approx {3\over 2} \theta _0 \sqrt {\eta {{f(1)}\over {f(\eta )}}}
(1-\Delta \xi _{_{\sc SG}}),
\label{theta_min_gamma}
\ee 
\be
\theta _{{\rm max}, \gamma }^{\sc SG} \approx {3\over 2} \theta _0 \sqrt {\eta {{f(1)}\over {f(\eta )}}}
~~~~{\rm and}~~~~\eta \approx \eta _{_{\gamma }} .
\label{theta_max_gamma}
\ee 
Thus, from eq. (\ref{omega1}) we can get
\be
\Omega _{\gamma } \approx {9\over 2}\pi \theta _0^2 \eta {{f(1)}\over {f(\eta )}}\Delta \xi _{_{\sc SG}}~~~~{\rm ster},
~~~~~\eta \approx \eta _{_{\gamma }} .
\label{omega2}
\ee
Here $\eta _{\gamma }$ is the radial distance of the high-energy emission 
site in the SG. 
Using expressions (\ref{nu}), (\ref{Lprim}), (\ref{omega2}), and the estimated values of $\Delta \xi _{_{\sc SG}}$ 
(see eqs [\ref{deltaxi1}] and [\ref{deltaxi2}]), we 
get

\noindent
{\it The case of $B \lsim 0.1~B_{\rm cr}$}

\be
L_{\gamma }(\Omega _{\gamma }) = {{\varepsilon _{\gamma }~
L_{\rm prim}}\over {\Omega _{\gamma}}} = 
3\times 10^{34} ~\varepsilon _{\gamma }~ 
L_{\rm sd, 35}^{3/7} P_{0.1}^{5/7} R_6^{17/7} 
\Lambda (\eta _{_{\gamma }})
~~~~~{\rm erg \cdot s^{-1} \cdot ster ^{-1}} .
\label{L/Omega1}
\ee

\noindent 
{\it The case of $B \gsim 0.1~B_{\rm cr}$}

\be
L_{\gamma }(\Omega _{\gamma }) = {{\varepsilon _{\gamma }~
L_{\rm prim}}\over {\Omega _{\gamma }}} = 
6 \times 10^{33} ~\varepsilon _{\gamma }~
L_{\rm sd, 35}^{4/7} P_{0.1}^{9/7}~R_6^{11/7} 
\Lambda (\eta _{_{\gamma }})~~~~~
{\rm erg \cdot s^{-1} \cdot ster ^{-1}} ,
\label{L/Omega2}
\ee
where 
\begin{eqnarray}
\Lambda (\eta ) & = & {{f(\eta )}\over \eta }\lambda ^{-8/7} I_{45}^{-6/7}    
\left\{ \kappa (1-\kappa )\left( 1- {1\over \eta ^3} \right) 
\cos ^2 \chi + \right. \nonumber \\
& &  \left. {9\over 8} \theta _0^2H^2(1)
\left[ {{H(\eta )}\over {H(1)}}\sqrt{\eta {{f(1)}\over {f(\eta )}}}-1 \right] 
\sin ^2 \chi \right\}, 
\end{eqnarray}
and $L_{\rm sd, 35} = L_{\rm sd}/10^{35}~{\rm erg\cdot s^{-1}}$.
In the above expressions the parameter $\varepsilon _{\gamma } $ ($0<\varepsilon _{\gamma } < 1$) is the efficiency of 
transforming the primary energetics into the energetics of high-energy  emission.
Note that in the above expressions $L_{\gamma  }(\Omega _{\gamma }) \propto L_{\rm sd}^{3/7}$ (low-B case) and  
$\propto L_{\rm sd}^{5/7}$ (high-B case) simply because $L_{\rm prim}\propto L_{\rm sd}^{1/7}$ and $\Omega _{\gamma } \propto L_{\rm sd}^{-2/7}$ (low-B case),  and $L_{\rm prim}\propto L_{\rm sd}^{5/14}$ and $\Omega _{\gamma } 
\propto L_{\rm sd}^{-3/14}$ (high-B case), respectively.  

The above expressions for $L_{\gamma }(\Omega _{\gamma })$ are equivalent to the observed quantity 
$\Phi _{\gamma }~d^2$, where $\Phi _{\gamma }$ is the high-energy bolometric flux observed at the Earth, and $d$ is the distance to the pulsar. Thus, the only free parameters 
(which may be determined by pair cascade simulations) are the efficiency 
$\varepsilon _{\gamma }$ and the parameter $\lambda$ determining the SG width. 
All others are intrinsic properties of the NS. Figure 3 shows the pulsar empirical 
(solid circles with error bars) and theoretical (upside-down triangles) values of $\Phi _{\gamma }d^2$ as a function 
of spin-down luminosity, $L_{\rm sd}$. The theoretical values are calculated for the parameters $\varepsilon _{\gamma } = 0.3$ and $\lambda = 0.1$ (see eq. [\ref{L/Omega1}]). Note that parameter $\varepsilon _{\gamma }$ can 
range from 0.2 to 0.5 in cascade calculation, and $\eta _{\gamma } = 3$. 
In Figure 3 the dashed line represents the absolute upper 
limit, where the spin-down luminosity is radiated into the unit solid angle, i.e. where 
$\Phi _{\gamma }d^2 = L_{\rm sd} / 1~{\rm ster}$. One can see that there is a good agreement for most
high-energy pulsars except several of the pulsars, Geminga and PSR B0656+14, having low $L_{\rm sd}$, and for J0218+4232, which is a millisecond pulsar. These pulsars are near or below the 
CR death line (see HM02, and next Section), and therefore have either very wide SGs or no SGs at all. 
All other high-energy pulsars depicted in Figure 3 
are above their CR death lines and are expected to have SGs.  

\subsection{Energetics of Returning Positrons in Slot Gap}

Let us estimate the maximum possible heating luminosity produced 
by returning positrons within the SG, provided that the PFF 
establishes at $\eta = \eta _{_{\sc PFF}}$. 

The general equation for the power of returning positrons can be written 
as (see e.g. MH97, eqs.[76]-[78]; and HM01, eq. [61])

\be 
L_{_+} = \alpha c \int _{_{{\cal S}_{_{\sc SG}}(\eta _{_{\sc PFF}})}} 
x_{_+} |\rho _{_{\sc GJ}}| \Phi d S,
\label{L+}
\ee 
where the integration is over the surface cut in a sphere at the 
radial distance $\eta _{_{\sc PFF}}$ by the SG boundaries, and 
$x_{_+} = \rho _{_+}(\eta _{_{\sc PFF}})/|\rho _{_{\sc GJ}}(\eta _{_{\sc PFF}})|$ is 
the charge density of returning positrons, scaled by the 
general-relativistic GJ charge density (see eq. [\ref{rhoGJ}]).

Let us calculate, for the sake of illustration, the fractional density of returning positrons 
for $\chi \approx 0$ assuming that $\rho _{_+} \approx |\rho _{_{\sc GJ}}(\eta _{_{\sc PFF}}) - \rho |/2$. 
Using formulae (\ref{rhoGJ}), (\ref{rho}) and representation of $\kappa $  
as $\kappa \approx 0.15~I_{45}/R_6^3$ (see \S~2.1, right after eq. [\ref{rho}]), we get 
\be
x_{_+} \sim {1\over 2} {{\kappa }
\over {(1-\kappa / \eta _{_{\sc PFF}}^3})}
\left( 1 - 1/\eta _{_{\sc PFF}}^3 \right) 
\approx {1\over 2} \kappa \approx 0.08~{{I_{45}}\over {R_6^3}}~~~{\rm for }~~\eta _{_{\sc PFF}} \gsim 2, 
\label{x+}
\ee
which means that for a nearly aligned pulsar the maximum fraction 
of positrons returning from the elevated pair fronts of SGs is independent 
of the PFF altitude, in contrast to the case of the PFFs establishing at low 
altitudes (see HM01, HM02) over the bulk core part of the PC. 

By inserting $\Phi (\eta , \theta , \phi_{\rm pc})$ and 
$\rho _{_{\sc GJ}}(\eta )$ into eq. (\ref{L+}) and 
performing the integration over $\theta $, in this case (cf. eqs.[\ref{theta_min_gamma}], 
[\ref{theta_max_gamma}]),\\  
from 
\be
\theta _{\rm min}^{\sc SG} = \theta _0 
\sqrt {\eta _{_{\sc PFF}}f(1)/f(\eta _{_{\sc PFF}})} 
(1-\Delta \xi _{_{\sc SG}})
\label{theta_min}
\ee
to 
\be
\theta _{\rm max}^{\sc SG} = \theta _0 
\sqrt {\eta _{_{\sc PFF}}f(1)/f(\eta _{_{\sc PFF}})},
\label{theta_max}
\ee  
we get
\be
L_{_+} = {1\over 4} \kappa ^2 (\Delta \xi _{_{\sc SG}} )^3 L _{\rm sd}.
\label{L+2}
\ee

\noindent
{\it The case of $B \lsim 0.1~B_{\rm cr}$}

\be
L_{_+} \approx 2 \times 10^{30} 
\lambda ^{-12/7} P_{0.1}^{-3/7}I_{45}^{5/7}
R_6^{-6/7} L_{\rm sd, 35}^{1/7}
~~~~~{\rm erg \cdot s^{-1}}  .
\label{L+LB}
\ee

\noindent
{\it The case $B \gsim 0.1~B_{\rm cr}$}

\be
L_{_+} \approx 4 \times  10^{29} 
\lambda ^{-12/7} P_{0.1}^{3/7} I_{45}^{5/7}
R_6^{-15/7} L_{\rm sd, 35}^{5/14}
~~~~~{\rm erg \cdot s^{-1}} .
\label{L+HB}
\ee

The contribution from heating by positrons returning from high 
altitudes of SGs should be taken into account in the estimates 
of the PC surface temperatures (see HM01, HM02)

It is interesting to note, that the condition of turning-on of the SG  acceleration in pulsars should be tightly related to the pair death line 
condition determined by the pair-formation above the central part of 
the PC (see e.g. HM02 for detailed discussion), simply because the very existence of the gap is determined by 
establishing the pair plasma column and the screening of $E_{||}$ by pairs. 
The analytic pair CR death line (see eq. [12] in HMZ02) reads 
\be 
\log {\dot P} \geq 
\left\{ \begin{array}{ll}
    {21\over 8}~\log P - {7\over 4}\log {\bar f}_{\rm prim}^{\rm min} - \Delta _{\rm I}^{\rm (CR)}(R,I)-14.6 & ,~~P\lsim P_{\ast }^{\rm (CR)}~~, \\
    {5\over 2}~\log P - 2\log {\bar f}_{\rm prim}^{\rm min} - \Delta _{\rm II}^{\rm (CR)}(R,I)-15.4 & ,~~P\gsim P_{\ast }^{\rm (CR)}~~,
\end{array} 
\right.
\label{DL_CR}
\ee
where 
\be
\Delta _{\rm I}^{\rm (CR)}(R,I) = 1.5(\log I_{45}-6.3\log R_6),
\ee
and 
\be
\Delta _{\rm II}^{\rm (CR)}(R,I) = 1.5(\log I_{45}-7.3\log R_6)
\ee
are corrections for the NS equation of state, and 
\be
P_{\ast }^{(CR)} = 0.1~B_{12}^{4/9}
\ee
is the value of pulsar spin period separating the regime of 
acceleration of primaries where the $E_{\parallel }$ is increasing 
with altitude (unsaturated)  from the regime where $E_{\parallel }$ remains practically constant (saturates).
In eq. (\ref{DL_CR}) ${\bar f}_{\rm prim }^{\rm min}$ is the minimum efficiency of converting spin-down luminosity into the luminosity of the primary beam needed for pair formation (see HM02 and HMZ02 for details). According to our calculation 
performed in HM02 the value of parameter ${\bar f}_{\rm prim}^{\rm min}$ ranges from 0.2 to 0.5. For the  
parameters of a canonical NS and for ${\bar f}_{\rm prim}^{\rm min} = 0.2$, we get 
\be
\log {\dot P} \gsim 2.5 \log P -14, 
\ee
as the required condition for SGs to form.

Note also that the relationship between the pulsar death line and the regime of SG 
acceleration may be justified only for the CR-produced pair front. 
Even though the SGs may still be formed in the case of the ICS pair production above the bulk of the PC, the density of pair plasma may not 
be sufficient to guarantee the efficiency of the geometrical screening. 
In this case the effective parameter $\Delta \nu _{_{\sc SG}}$ 
may significantly differ from that calculated in this paper, including 
the case where it approaches 1 (no geometrical screening within the 
SG). Our previous calculation of the ICS PFFs indicates (see e.g. HM01, HM02) that this situation may well occur in some pulsars that are slightly below the death line (e.g. in millisecond and relatively old pulsars). In this paper we do not intend to discuss these 
non-ideal SGs.   

\section{Pair Cascade Simulation}

Before we discuss our numerical simulation it is worth mentioning that 
there is a long history associated with the ``prototype" of a SG model, 
the pure geometrical model of pulsar emission known as the hollow cone model. 
By analyzing the smooth variation of the direction of linear polarization
within the radio pulsar pulse window Radhakrishnan \& Cooke (1969) and 
Radhakrishnan (1969) realized that these observations suggested that 
pulsars must be magnetized NSs which produced relativistic 
particles from near the magnetic PCs. Komesaroff (1970) showed that if 
the radio emission is due to CR then the pulsar beam must be a hollow cone. 
The double pulses observed in some pulsars, the pulse width, polarization, 
as well as the spectrum of pulses found a natural explanation within 
this model.

In order to explore the characteristics of high-energy emission produced by particles 
accelerated in the SG, we have carried out Monte Carlo simulations of the pair cascades
initiated by primary electrons accelerated by the $E_{\parallel}$ in the gap.  This 
calculation is very similar to the pair cascade simulation described by DH96.  In that paper,
primary electrons were accelerated above a pulsar PC by a simple model $E_{\parallel}$
described by starting and terminating altitudes above the NS surface, a constant term and a 
linear term.  These parameters were adjusted to produce both a sufficient final Lorentz factor
of the primary electrons and a wide enough gamma-ray beam geometry to explain the EGRET
spectrum and pulse profile of the Vela pulsar.  DH96 found that a $E_{\parallel}$ starting
altitude of 2 stellar radii was required, but no physical justification was provided to support
this value.  DH96 also artificially enhanced the primary particle flux at the PC rim in order to 
produce the hollow emission cone, narrow pulses and low level of bridge emission necessary to 
model the Vela profile.

In the present study, we use the SG $E_{\parallel} = -\nabla\Phi$ derived from 
the potential in eq. (\ref{Phi}).  The acceleration rate as a function of distance is
\be 
{{d \gamma }\over {ds}} = {{eE_{\parallel}}\over {m_{\rm e}c^2}} 
\approx 54~[1-{3\over 8}(1+7\delta)]~{{R_6^2~B_{12}^{\delta }}\over {(\lambda ^{4}I_{45}^{3})^{2/7}}}
\left( {\kappa\over \eta^4}\cos\chi + {1\over 4}\theta_0\eta^{-1/2}\sin\chi
\cos\phi_{\rm pc}\right) (1-\xi _{\ast }^2), 
\label{Epar}
\ee
where $\delta $ is defined right after eq. (\ref{Phi_last}).
In this formula, for the sake of illustration,  we pull out the general-relativistic correction from the second term to get its simple explicit $\eta $-dependence. Even though in most cases the second term in the above equation is less 
than the first one, in our numerical calculations we use the exact formula  
for $E_{\parallel }$ based on expression (\ref{Phi}).    
The first term in parentheses is the contribution to acceleration from the inertial frame-dragging 
effect, where $\kappa \approx 0.15~I_{45}/R_6^3$ is the general relativistic parameter that depends 
on the NS compactness.  The second term is the contribution from the field-line curvature, similar 
to the accelerating field of AS79.  Factors of order one have been omitted
from this term in order to simplify the expression in the potential of eq. (\ref{Phi}). 
Note that the SG acceleration rate given above is only mildly dependent on B and P. 
We assume that primary electrons begin their acceleration at the NS surface on a central field line of the SG, $\xi_* = 0$.  
We follow the electrons as they gain energy and follow
their CR photons through the local magnetic field, accumulating the probability
for creating an electron-positron pair.  If the CR photons pair produce, the synchrotron emission 
of the created pair is simulated and the synchrotron radiation photons are followed in the same way to determine if
they pair produce or escape.  Details of the cascade simulation may be found in DH96 and
Harding et al. (1997).  We ignore inverse Compton emission of the primaries and pairs and 
photon splitting of the CR photons.  

To illustrate the features of the SG acceleration and cascades, we use the period 
($P_{0.1} = 0.33$) and surface magnetic field ($B_{12} = 5$) of the Crab pulsar for the case $B < 0.1~B_{\rm cr}$ in eq. (\ref{Epar}). 
A stellar radius of 16 km is assumed. Although the SG is a region devoid of pair plasma 
near the surface, full pair cascades can develop above the PFFs that form at high altitudes
along field lines near the PC rim.  In contrast to the pair cascades that form in the core
of the PC and terminate within a stellar radius of the NS surface, the pair cascades from
particles accelerated in the SG develop over a larger distance and the screening of
the electric field by the pairs thus requires a much larger distance.  The primary electrons
reach Lorentz factors of $10^7$ in the SG before becoming radiation-reaction limited.  
However, the pair cascades begin at lower Lorentz factors but the pair 
multiplicity builds up more slowly in comparison to pair cascades in the PC core (see HM01), so 
that the screening distance in the SG is relatively large and can be one or two stellar 
radii.
We find that the pair cascade starts at an altitude of 0.5 - 1.0 stellar radii and continues
up to 3-4 radii.  Since the bulk of the
escaping high-energy photons are produced between 2 and 4 NS radii, the opening angle of the high-energy beam is much larger (about 20 degrees for the
Crab period) than for near-surface PC emission.  

Figure 4 shows plots of the distribution on the sky of escaping cascade radiation above 100 keV
as a function 
of phase, $\phi$, and viewing angle with respect to the rotation axis, $\zeta$, for different  assumed values 
of the inclination.  For relatively small inclination angles that are comparable to the high-
energy beam opening angle, emission can be   seen over a large range of observer angles.  As the 
inclination angle decreases, the hollow cone pattern becomes smaller and emission is visible over 
a smaller range of viewing angles.  Also at smaller inclination angles, the emission hollow cone 
is complete because particle acceleration takes place at all phase angles around the magnetic 
pole.  This is a result of the fact that the frame dragging term in the potential is independent 
of $\phi_{\rm pc}$.  At larger inclination angles, 
our results show that the emission occurs over only part of the hollow cone.  This is due to the 
$\sin\chi\cos\phi_{\rm pc}$ dependence of the second term in the potential 
due to the ``favorably curved" 
field line acceleration that was the only source of accelerated particle in Arons' potential.   
At large inclination angles this part of the potential dominates as the frame dragging term is 
declining.  Because the favorably curved field lines from the two poles accelerate particles in 
opposite hemispheres, it is impossible for a single observer to view both poles of the pulsar in 
this model unless the high-energy beam opening angle is unrealistically large. Instead, at 
large viewing angles an observer would see a single narrow pulse.  

Pulse profiles are formed by slicing across the phase plots in Figure 4 at a constant viewing 
angle.  Examples of pulse profiles at several different inclination and viewing angles are shown 
in Figure 5.   It is evident that widely spaced double-peaked profiles can result at relatively 
small inclination by slicing through the middle of the hollow cone.  Profiles with two pulses 
separated by more than 100 degrees, as are seen from the Crab,  Vela and Geminga, can be formed 
when the inclination angle is comparable to the beam opening angle (about 20 degrees in this model 
for the Crab).   Broad single pulses, as seen from PSR B1509-58, may be produced when the 
observer cuts along the edge of the cone.   As the inclination angle increases, the maximum pulse 
separation decreases and the pulses themselves also become narrower.  Bridge emission as seen in 
Vela and Geminga high-energy profiles (and in the Crab profile around 1 MeV) is not produced in
this calculation since we have modeled only cascades and emission from the SG, which  
produces a pure hollow cone beam.  The core cascades will produce additional emission on field lines interior to the SG which will fill in the profile between the two pulses and produce 
bridge emission.  Emission from the full PC will be included in future modeling.  Emission can be 
seen in this model at all phases because primary electrons will continue to radiate curvature 
emission at higher altitudes (all the way to the light cylinder) above the termination of the 
pair cascade.  This is the ``off-beam" emission (Harding \& Zhang 2001) that will be much softer 
than the on-beam emission from the cone.  

Our cascade simulation includes aberration of the emitted photons as well as delay times between 
photons emitted at different altitudes.  Both of these effects become more important at the 
higher altitudes of the SG cascades and will produce asymmetries in the pulse profiles, 
especially for fast rotators.   For example, aberration will produce a pair attenuation cutoff 
at lower energies for the pulse at the leading edge of the profile because the photon emitted at 
the leading edge attain large enough angles to the magnetic field to produce a pair in a shorter 
distance than photons emitted at the trailing edge.  This effect was studied recently by Dyks \& 
Rudak (2002), who found that the observed disappearance of the first pulse in double-peaked 
profiles of the Crab, Vela and Geminga pulsars could be produced if the photons were emitted at 
altitudes of several stellar radii above the surface.  Emission from the SG  cascades will 
therefore naturally produce this effect.

\section{Discussion and Conclusions}

In this paper we revise the regime of electron acceleration 
within the SG regions above pulsar PCs. We have demonstrated that our  
treatment of acceleration within the SG allows us to address the longstanding 
issues of $\gamma $-ray production at high altitudes in the pulsar inner magnetosphere, 
interpretation of multifrequency light curves of pulsars, and emission energetics. 
The fundamental differences between the SG model presented 
in this paper and the original SG model of AS79 are 

\begin{enumerate}

\item Our model takes into account the effect of electric field screening 
within the gap caused by the narrowness of the space between the gap 
equipotential surfaces (effect of nearly parallel conducting boundaries on the 
accelerating electric field). 

\item Our model incorporates the effect of general-relativistic frame dragging, 
which in most cases determines the PC acceleration properties.
As a consequence of this, our SG model is operable for all azimuthal angles around 
the PC and for all pulsar obliquities, whereas the classical Arons-Scharlemann 
model operates only for the so-called ``favorably curved" (curved toward 
the rotation axis) field lines, and for nearly orthogonal rotators.

\item  We consider the high-energy emission from pair cascades at the inner edge 
of the SG while 
Arons (1983, 1996) considered only the CR from primaries in the SG. 

\end{enumerate}      

The first two effects make it possible to initially boost the electron 
acceleration and then prolong it up to very high altitudes (from few to 
several stellar radii) where they generate pair-producing CR photons. As 
a result of pair formation at high altitudes, the high-energy photons 
get emitted into a beam that is larger than the solid angle 
pertaining to the SG itself.  These effects also enable the functioning 
of SGs in all pulsars which are above the CR pair-formation death lines 
(see HMZ02 for the details of our most recent death-line calculations). 
The second property significantly enlarges the allowed parameter space for 
the pulsar obliquity and viewing angles, in particular to small inclination
and nearly aligned rotators where the original SG model failed to produce 
any acceleration.  Within the SG 
model proposed in this paper both double and single high-energy pulses 
are possible, and in the case of double pulses the separation between the 
peaks may be anywhere from near zero to 180 degrees. The asymmetry of 
double-peaked high-energy pulses (one peak is higher than other), and 
misalignment of $\gamma $-ray light curves with the X-ray light curves 
may be attributed to the effect of photon aberration which 
becomes important for high-energy photons generated at high altitudes.
Some other effects, such as e.g. effect of NS equation-of-state and effects of very 
strong stellar magnetic field will be discussed separately in our subsequent 
publications. Also, we plan to investigate the details of formation of 
high-energy spectra in SGs of pulsars and perform comparison with the 
observational data. 

In this paper we calculated the predicted high-energy luminosity per 
unit solid angle from the SG of pulsars as a function of pulsar spin-down 
luminosity. Our theoretical calculations agree well with the observational 
values for $\gamma $-ray pulsars. Given the fact that our model has only 
two major uncertainties, such as the uncertainty in efficiency of 
conversion of primary 
energetics into the energetics of $\gamma $-ray emission and uncertainty 
of our calculation of SG width, the agreement is rather remarkable. Note 
that the uncertainty of our SG width calculation translates into the 
uncertainty of order of a few for our calculation of high-energy luminosity 
per unit solid angle which is comparable to a typical uncertainty of 
the observations. Our calculations depicted in Figure 3 were performed 
for the efficiency of conversion of primary energetics into the energetics 
of high-energy emission $\varepsilon _{\gamma }=0.3$, which does not seem to be unreasonable.  

The SG cascades will produce a component of electron-positron pairs at
high altitudes, in addition to and distinct from the pair component produced
by the low-altitude core cascades.  One might speculate that these two pair components
could be the origin of the core and conal radio emission components that have been
identified from radio morphology studies (e.g. Rankin 1993, Lyne \& Manchester 1988). 
It will be interesting to explore the properties of these two pair components, which
may have different characteristics.

We presented our illustrative calculations of $\gamma $-ray 
light curves and double-peaked pulse profiles for the parameters 
of the Crab pulsar. We believe that our simulated high-energy light curves 
and pulse profiles can be used to constrain the obliquity and viewing 
angles for $\gamma $-ray pulsars. We plan to perform detailed and 
extensive simulations of this kind in the future. We also plan to 
perform comparative analysis of two-pole and one-pole models for 
selected pulsars with available multifrequency light curves and 
indirect studies of their viewing and obliquity angles.

\acknowledgments 
We would like to thank Matthew Baring and Bing Zhang for discussions and many helpful comments on the manuscript.  We also acknowledge support from the NASA Astrophysics Theory Program.

\newpage

~
\vskip -1.0in
\figureout{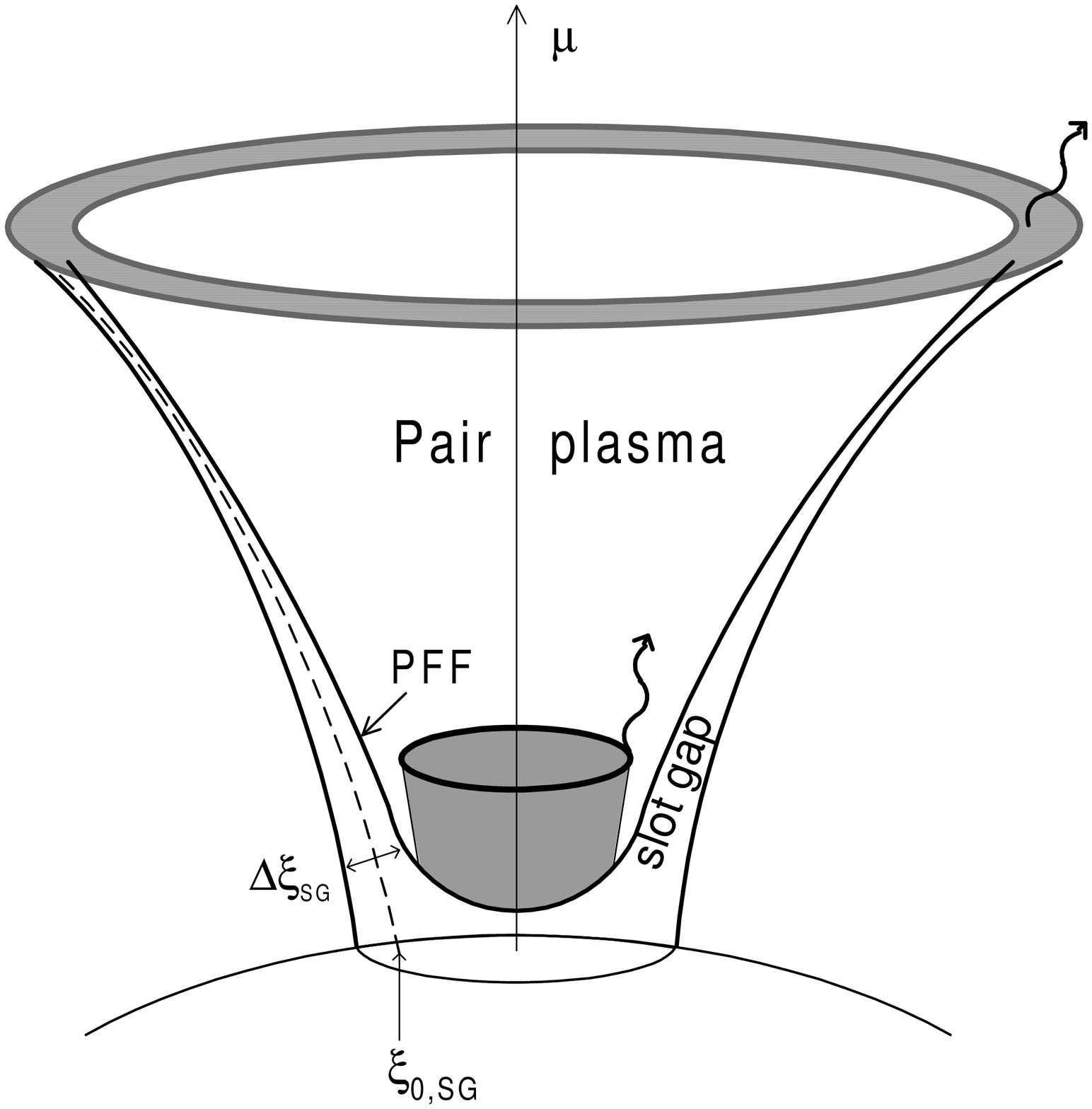}{0}{
Schematic illustration of polar cap geometry, showing the outer boundary of the open field line 
region (where $E_\parallel = 0$) and the curved shape of the pair formation front (PFF) which 
asymptotically approaches the boundary at high altitude.  The slot gap  exists between the pair plasma which results from the pair cascades above the PFF and the outer boundary.  A narrow
beam of high-energy emission originates from the low-altitude cascade on field lines interior 
to the slot gap.  A broader, hollow-cone beam originates from the high-altitude cascade above
the interior edge of the slot gap. $\Delta \xi$ is the slot gap thickness (see text) and 
$\theta_{0, SG}$ is the colatitude at the center of the slot gap.
}    

\figureout{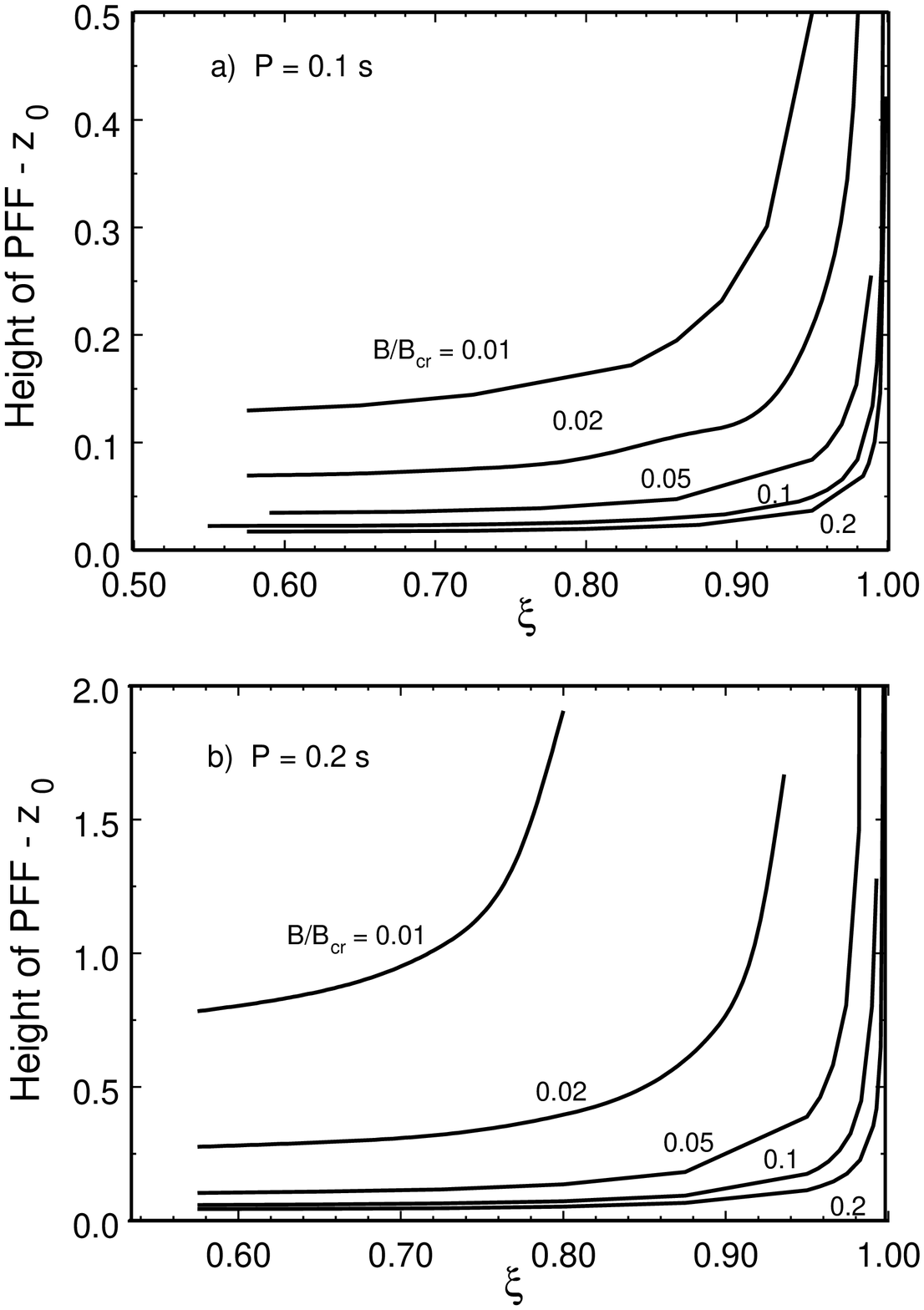}{0}{
Height of the pair formation front for curvature radiation photons above the neutron star (NS) surface $z_0$, in units of NS radius R,
as a funtion of magnetic colatitude in units of polar cap half-angle, $\xi \equiv \theta/\theta_0$,
for different values of surface magnetic field strength B, in units of critical field 
$B_{\rm cr}$ and $\chi = 30^o$.
    }    

\psfig{figure=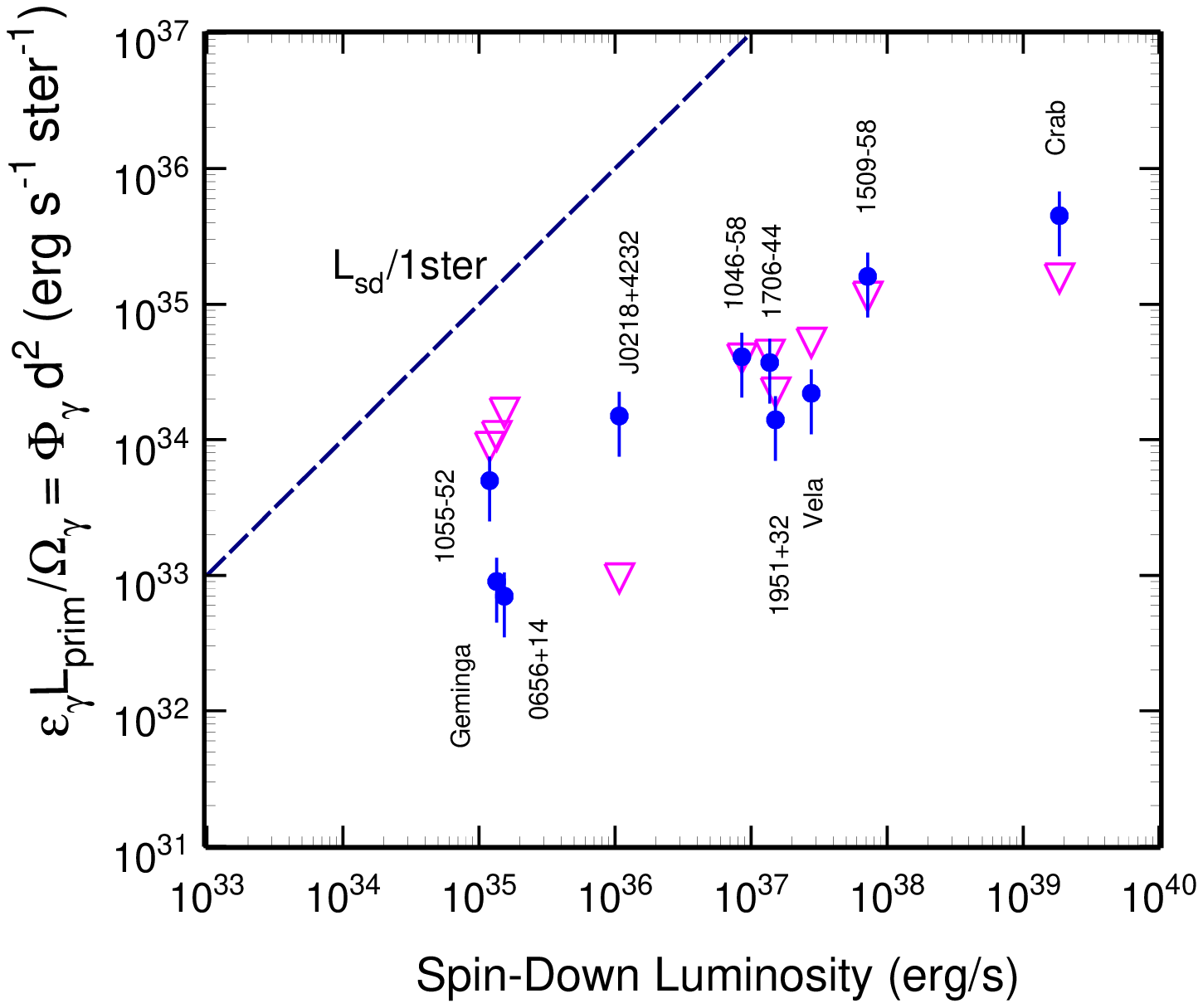,width=7.0in,angle=0}
\figcaption{
Observed flux above 1 keV, $\Phi_{\gamma}$, times distance squared (from Thompson (2001))
(solid circles) and theoretical values of specific high-energy luminosity from the slot gap, $\varepsilon _{\gamma }~L_{\rm prim}/ \Omega _{\gamma}$ from eq. (\ref{L/Omega1}) 
(upside-down triangles) vs. spin-down luminosity for known $\gamma$-ray pulsars.  An
efficiency of $\varepsilon_{\gamma} = 0.3$ was assumed. Also $\lambda = 0.1$, $\eta_{\gamma} = 3$ 
and the stellar parameters $R_6=1.6$ and $I_{45}=4$ were used. 
    }    
\newpage
~
\hskip -0.5in
\psfig{figure=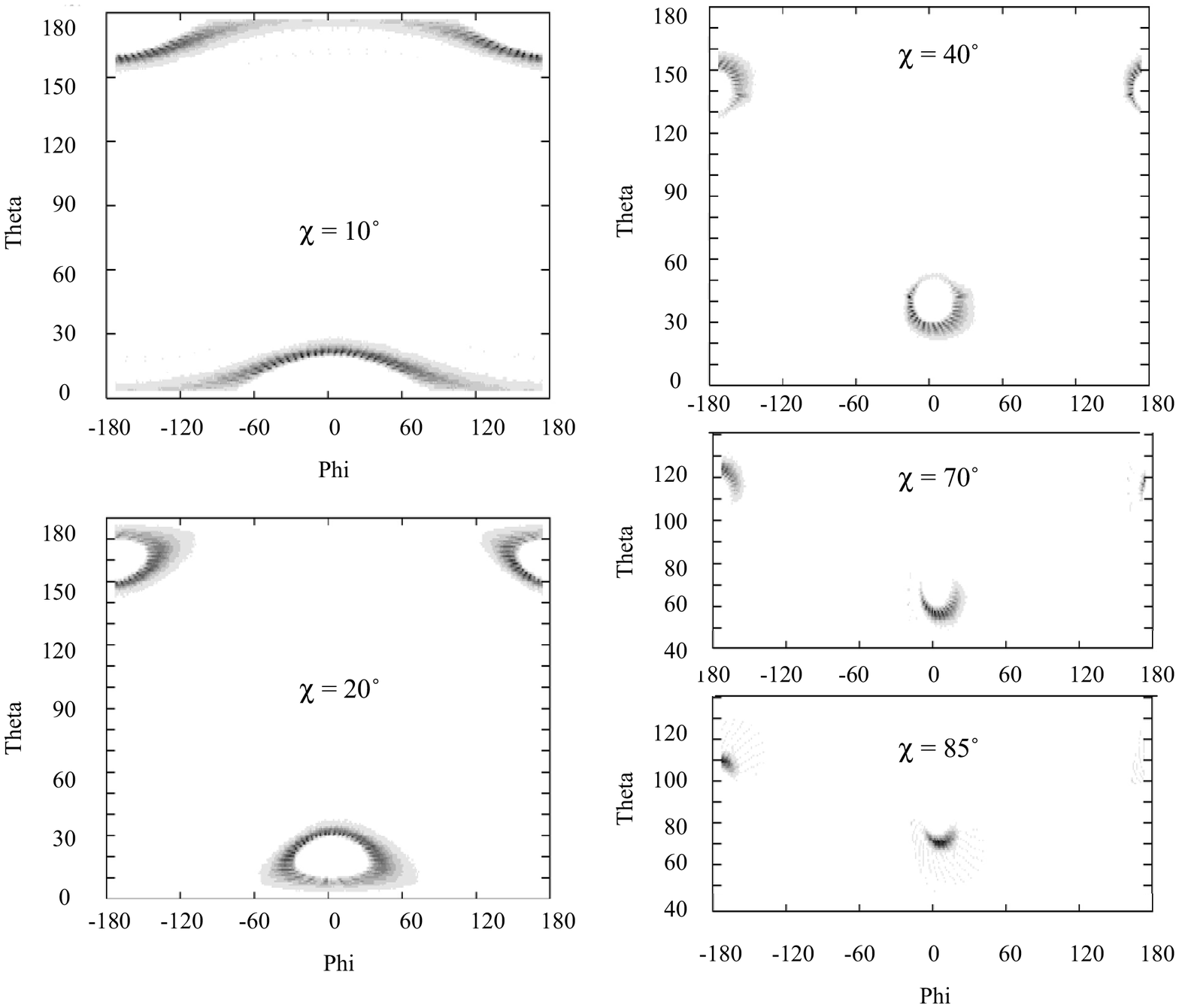,width=7.5in,angle=0}
\figcaption{
Angular distribution of photon emission above 100 keV from pair cascades above the slot gap
(using a linear 10-level gray scale), for different values of inclination angle $\chi$.
Theta and phi are polar and azimuthal angles with respect to the pulsar rotation axis.  Cascade simulations assumed values, $P = 33$ ms, 
$B = 5\times 10^{12}$ G, of the Crab pulsar.
Extended faint tracks are high-altitude curvature radiation from primary electrons above
the cascade region and are an artifact of the resolution in magnetic phase angle $\phi_{\rm pc}$
around the polar cap.
    }    
\newpage
~
\hskip -0.5in
\psfig{figure=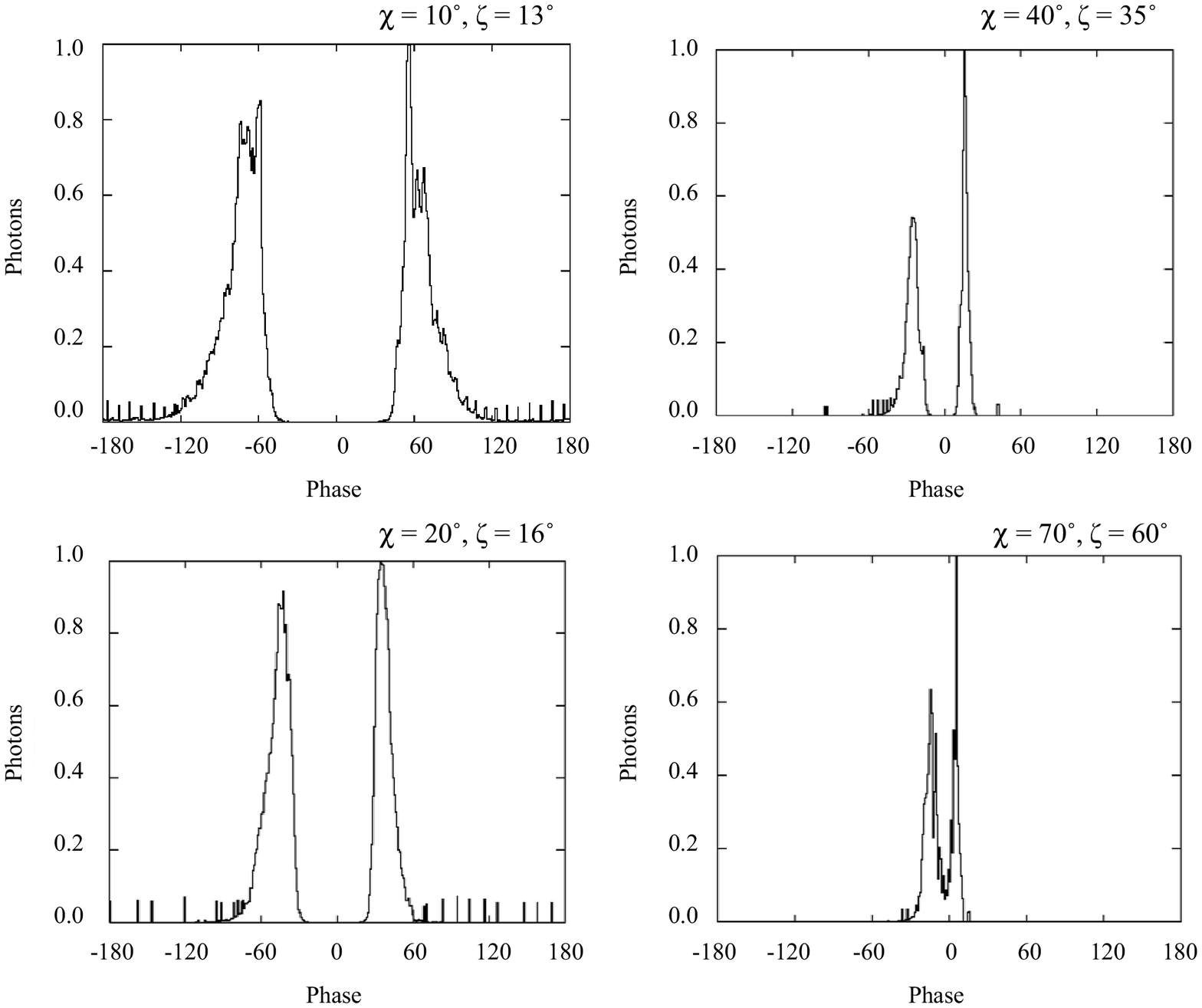,width=8.0in,angle=0}
\figcaption{
Theoretical pulse profiles of photon emission above 100 keV from the slot gap pair cascade 
at various viewing angles, $\zeta$, generated by making cuts through the angular distributions shown in Figure 4 at constant phase angle theta.  The spikes on the leading and trailing edges
of the peaks in some profiles are artifacts of the resolution in magnetic phase angle 
$\phi_{\rm pc}$ around the polar cap.
    }    

\end{document}